\documentclass[conference]{IEEEtran}
\IEEEoverridecommandlockouts
\usepackage{cite}
\usepackage{amsmath,amssymb,amsfonts}
\usepackage{algorithmic}
\usepackage{graphicx}
\usepackage{textcomp}
\usepackage{xcolor}
\usepackage{pgfplots}
\pgfplotsset{compat=1.15}
\usepgfplotslibrary{statistics}
\usetikzlibrary{patterns}
\usepackage{bm}
\usepackage{tabularx}
\usepackage{subcaption}
\usepackage{balance}

\usepackage{mathtools}	
\usepackage{amssymb}
\usepackage{amsthm}

\usepackage[utf8]{inputenc}

\usepackage{ifthen}

\usepackage{microtype}

\usepackage{caption}

\usepackage{bm}

\usepackage[draft,nomargin,marginclue,inline]{fixme}
\makeatletter
\renewcommand*\FXLayoutMarginClue[3]{%
	\marginpar[%
	\raggedleft\@fxuseface{margin}\textcolor{red}{\ignorespaces $ \Rightarrow $}]{%
		\raggedright\@fxuseface{margin}\textcolor{red}{\ignorespaces $ \Leftarrow $}}}
\makeatother	

\usepackage{tumcolor}

\usepackage{pgfplotstable}	

\pgfplotsset{
	discard if/.style 2 args={
		x filter/.append code={
			\edef\tempa{\thisrow{#1}}
			\edef\tempb{#2}
			\ifx\tempa\tempb
			
			\fi
		}
	},
	discard if not/.style 2 args={
		x filter/.append code={
			\edef\tempa{\thisrow{#1}}
			\edef\tempb{#2}
			\ifx\tempa\tempb
			\else
			
			\fi
		}
	}
}

\usepackage{cite}
\usepackage{hyperref}
\hypersetup{
	colorlinks = true,
	linkbordercolor = {white},
	citecolor = {black},
	linkcolor = {black},
	urlcolor = {black}
}
\makeatletter
\renewcommand*{\eqref}[1]{%
	\hyperref[{#1}]{\textup{\tagform@{\ref*{#1}}}}%
}
\makeatother

\usepackage[acronym,shortcuts]{glossaries}
\newacronym{blmmse}{BLMMSE}{Bussgang LMMSE}
\newacronym{cnn}{CNN}{convolutional neural network}
\newacronym{dft}{DFT}{discrete Fourier transform}
\newacronym{em}{EM}{expectation-maximization}
\newacronym{emiht}{EM-IHT}{EM algorithm with IHT}
\newacronym{iht}{IHT}{iterative hard thresholding}
\newacronym{lmmse}{LMMSE}{linear minimum mean square error}
\newacronym{mimo}{MIMO}{multiple-input multiple-output}
\newacronym{mmse}{MMSE}{minimum mean square error}
\newacronym{mse}{MSE}{mean square error}
\newacronym{relu}{ReLU}{rectified linear unit}
\newacronym{snr}{SNR}{signal-to-noise ratio}
\newacronym{adc}{ADC}{analog-to-digital converter}
\newacronym{nn}{NN}{neural network}
\newacronym{re}{RE}{resource element}

\newcommand{\op}[1]{{\operatorname{#1}}}
\newcommand{\uproman}[1]{\uppercase\expandafter{\romannumeral#1}}

\newcommand{\vect}{\operatorname{vec}}
\newcommand{\diag}{\operatorname{diag}}

\newcommand{\eye}{\bm{\op{I}}}
\newcommand{\B}[1]{\bm{#1}}
\newcommand{\inv}{^{-1}}

\newcommand{\h}{^{\op H}}
\newcommand{\T}{^{\op T}}

\newcommand{\dB}{$\operatorname{dB}$}

\usepackage{cleveref}

\usepackage{enumitem}

\newcommand{\lineWidth}{1.0pt}
\tikzset{algorithm1/.style={mark options={solid},color=TUMBeamerBlue, line width=\lineWidth, mark=square, dashed}}

\newcommand*{\C}{\mathbb{C}}

\newcommand*{\R}{\mathbb{R}}

\newlength{\leftstackrelawd}
\newlength{\leftstackrelbwd}
\def\leftstackrel#1#2{\settowidth{\leftstackrelawd}%
	{${{}^{#1}}$}\settowidth{\leftstackrelbwd}{$#2$}%
	\addtolength{\leftstackrelawd}{-\leftstackrelbwd}%
	\leavevmode\ifthenelse{\lengthtest{\leftstackrelawd>0pt}}%
	{\kern-.5\leftstackrelawd}{}\mathrel{\mathop{#2}\limits^{#1}}}

\newcommand{\mbA}{\bm{A}}

\newcommand{\mbH}{\bm{H}}

\newcommand{\mbh}{\bm{h}}

\newcommand{\calN}{\mathcal{N}}

\newacronym{gmm}{GMM}{Gaussian mixture model}
\newacronym{pdf}{PDF}{probability density function}
\newacronym{cme}{CME}{conditional mean estimator}
\newacronym{siso}{SISO}{single-input single-output}
\newacronym{pdp}{PDP}{power delay profile}
\newacronym{dps}{DS}{Doppler spectrum}
\newacronym{fft}{FFT}{fast Fourier transform}
\newacronym{ofdm}{OFDM}{orthogonal frequency-division multiplexing}
\newacronym{ml}{ML}{machine learning}
\newacronym{mt}{MT}{mobile terminal}
\newacronym{bs}{BS}{base station}
\newacronym{ul}{UL}{uplink}
\newacronym{dl}{DL}{downlink}
\newacronym{fdd}{FDD}{frequency division duplex}
\newacronym{tdd}{TDD}{time division duplex}
\newacronym{ce}{CE}{channel estimation}
\newacronym{los}{LOS}{line of sight}
\newacronym{nlos}{NLOS}{non-line of sight}
\newacronym{o2i}{O2I}{outdoor-to-indoor}
\newacronym{uma}{UMa}{urban macrocell}

\def\BibTeX{{\rm B\kern-.05em{\sc i\kern-.025em b}\kern-.08em
    T\kern-.1667em\lower.7ex\hbox{E}\kern-.125emX}}

\pgfplotsset{tick label style={font=\small},label style={font=\small},legend style={font=\scriptsize}}

\definecolor{myblack}{RGB}{70,70,70}
\definecolor{myblue}{RGB}{65,105,225}
\definecolor{mygreen}{RGB}{0,139,139}
\definecolor{myorange}{RGB}{255,150,0}
\definecolor{myred}{RGB}{255,69,0}
\definecolor{mylila}{RGB}{153,50,204}

\newcommand{\legendgmm}           {GMM full}

\newcommand{\legendgmmkron}       {GMM kron}

\newcommand{\legendtwodgmmmid}          {GMM 2$\times$1D}
\newcommand{\legendtwodgmmmidcirc}         {GMM 2$\times$1D-circ}
\newcommand{\legendpdpkron}      {PDP+DS kron}
\newcommand{\legendpdptwod}      {PDP+DS 2$\times$1D}

\newcommand{\plotwidth}{1\columnwidth}
\newcommand{\plotheight}{0.6\columnwidth}

\newcommand{\quadriga}{QuaDRiGa }
\newcommand{\pdp}{\text{PDP} }
\newcommand{\dps}{\text{DS}}

\newcommand{\marksize}{1.6pt}

\tikzset{genie/.style={mark options={solid},color=black, line width=\lineWidth}}
\tikzset{global/.style={mark options={solid},color=TUMBeamerGreen, line width=\lineWidth, mark=triangle, mark size=\marksize, dashed}}
\tikzset{gmm/.style={mark options={solid},color=TUMBeamerBlue, line width=\lineWidth, mark=square, mark size=\marksize}}
\tikzset{gmm_diag/.style={mark options={solid},color=TUMBeamerLightBlue, line width=\lineWidth, mark=square, mark size=\marksize, dotted}}
\tikzset{gmm_kron/.style={mark options={solid},color=black, line width=\lineWidth, mark=square, mark size=\marksize, dotted}}
\tikzset{2x1gmm/.style={mark options={solid},color=mylila, line width=\lineWidth, mark=diamond, mark size=\marksize}}
\tikzset{2x1gmm_mid/.style={mark options={solid},color=TUMBeamerGreen, line width=\lineWidth, mark=triangle, mark size=\marksize, dashed}}
\tikzset{2x1gmm_mid_circ/.style={mark options={solid},color=TUMBeamerRed, line width=\lineWidth, mark=x, mark size=\marksize, dashdotted}}
\tikzset{old/.style={mark options={solid},color=TUMBeamerYellow, line width=\lineWidth, mark size=\marksize, dashdotted}}
\tikzset{em_gm_gamp/.style={mark options={solid},color=TUMBeamerRed, line width=\lineWidth, mark=triangle, mark size=\marksize, dashdotted}}
\tikzset{double_gmm_genie/.style={mark options={solid},color=gray, line width=\lineWidth, mark=triangle, mark size=\marksize, dashdotted}}
\tikzset{pdp2x1/.style={mark options={solid},color=gray, line width=\lineWidth, mark=o, mark size=\marksize, solid}}
\tikzset{pdp_kron/.style={mark options={solid},color=black, line width=\lineWidth, mark=o, mark size=\marksize, dotted}}
\tikzset{gmm_toep/.style={mark options={solid},color=TUMBeamerGreen, line width=\lineWidth, mark=x, mark size=\marksize, dashed}}

\begin{document}

\title{Channel Estimation based on Gaussian Mixture Models with Structured Covariances
\thanks{This work was funded by Huawei Sweden Technologies AB, Lund.}
}
\author{
	\centerline{Benedikt Fesl, Michael Joham, Sha Hu$^\dagger$, Michael Koller, Nurettin Turan, and Wolfgang Utschick}\\
	\IEEEauthorblockA{School of Computation, Information and Technology, Technical University of Munich, Germany\\
	$^\dagger$Huawei Sweden Technologies AB, Lund, Sweden\\
	Email: \{benedikt.fesl, joham, michael.koller, nurettin.turan, utschick\}@tum.de\\
	$^\dagger$Email: hu.sha@huawei.com
    }
}

\maketitle

\begin{abstract}
In this work, we propose variations of a Gaussian mixture model (GMM) based channel estimator that was recently proven to be asymptotically optimal in the minimum mean square error (MMSE) sense. We account for the need of low computational complexity in the online estimation and low cost for training and storage in practical applications. 
To this end, we discuss modifications of the underlying expectation-maximization (EM) algorithm, which is needed to fit the parameters of the GMM, to allow for structurally constrained covariances.
Further, we investigate splitting the 2D time and frequency estimation problem in wideband systems into cascaded 1D estimations with the help of the GMM. The proposed cascaded GMM approach drastically reduces the complexity and memory requirements.
We observe that due to the training on realistic channel data, the proposed GMM estimators seem to inherently perform a trade-off between saving complexity/parameters and estimation performance.
We compare these low-complexity approaches to a practical and low cost method that relies on the power delay profile (PDP) and the Doppler spectrum (DS). We argue that, with the training on scenario-specific data from the environment, these practical baselines are outperformed by far with equal estimation complexity.
\end{abstract}


\section{Introduction}
\begin{figure}[b]
\onecolumn
\centering
\copyright This work has been submitted to the IEEE for possible publication. Copyright may be transferred without notice, after which this version may no longer be accessible.

\vspace{-2.05cm}
\twocolumn
\end{figure}
In recent years, \ac{ml} has emerged as a powerful tool in order to meet the increasing requirements 
for \ac{ce} \cite{8052521}. 
\ac{ml} approaches for communications are designed as \textit{end-to-end learning}, e.g. \cite{8985539}, or \textit{model-aided learning}, e.g. \cite{8272484}, or a mixture of both.
The basis for all \ac{ml} methods is the construction of appropriate datasets for training. In this work, we focus on the idea that the dataset is tailored for a specific scenario/environment of a \ac{bs} cell. 
Such environments have immanent characteristics (e.g., buildings, streets) that can be seen as a prior for the communication tasks that are performed in this cell.
These characteristics can not be modeled, but \ac{ml} methods trained on appropriate data have access to these priors by learning their underlying distribution within the data during training.   

In \cite{Koller2021icassp,4536367}, a \ac{gmm} based channel estimator is introduced that is proven to be the asymptotically optimal \ac{cme} for an infinite number of mixture components. 
Since the \ac{cme} is approximated by the \ac{gmm} estimator that is fitted with training data from a specific \ac{bs} environment, this is a model-aided learning approach.
Although the channel estimator performs well also for a finite number of components, the evaluation of the \ac{gmm} estimator might lead to a prohibitively large complexity and memory overhead. 
However, by introducing certain assumptions on the system model, reasonable structural approximations of the covariance matrix can be found, e.g., a Toeplitz or circulant structure \cite{CIT-006}. In addition, the 2D estimation of a time-frequency channel can be decomposed by exploiting a Kronecker structure or by cascaded 1D estimation~\cite{598897}.

In contrast, practical approaches focus on low cost methods that depend on specific parameters that are simple to estimate, such as the \ac{pdp} and the \ac{dps} \cite{6082363,4012493}. Exploiting the known relations of the time/frequency correlations and the \ac{dps}/\ac{pdp} by the Fourier transform using the Wiener–Khinchin theorem \cite{89767565}, one can evaluate an \ac{lmmse} estimate with these parameters.

The contributions of this work are summarized as follows. 
We propose different low-complexity approaches of the \ac{gmm} based channel estimator from \cite{Koller2021icassp,4536367} for wideband channels. In particular, we constrain the respective covariance matrices of each component to circulant or Toeplitz matrices by \ac{dft} based decompositions. Especially, the Toeplitz case, realized by specific projections in the underlying \ac{em} algorithm, shows a strong performance close to the \ac{gmm} estimator having unconstrained covariances with much more parameters.
We then propose a 2$\times$1D estimator that performs cascaded 1D estimations of the time and frequency channels in the online phase. For this, we fit one \ac{gmm} for each dimension separately in the offline phase. This approach drastically reduces the complexity and storage requirements.
We compare the respective low cost approaches to an \ac{lmmse} estimator based on \ac{dps}/\ac{pdp}, which are computed with knowledge of the true instantaneous channel realization. 

\section{System and Channel Model}\label{sec:system_model}

\begin{figure}[t]
\centering
	\newcommand*{\xMin}{1}%
	\newcommand*{\xMax}{25}%
	\newcommand*{\yMin}{1}%
	\newcommand*{\yMax}{15}%
	\centering
	\begin{tikzpicture}[scale=0.2]
		\foreach \i in {\xMin,...,\xMax} {
			\draw [very thin,gray] (\i,\yMin) -- (\i,\yMax) ;
		}
		\foreach \i in {\yMin,...,\yMax} {
			\draw [very thin,gray] (\xMin,\i) -- (\xMax,\i);
		}
		\foreach \i in {1,5,10,15,20,24} {
			\node at (\i+0.5, 0.3) {\scriptsize$\i$};
		}
		\foreach \i in {1,5,10,14} {
			\node at (0.3, \i+0.5) {\scriptsize$\i$};
		}
		\foreach \i in {0, 2, 5, 7, 10, 12, 15, 17, 20, 23} {
		    \foreach \k in {0, 3, 6, 9, 13} {
			\draw[fill=TUMBeamerBlue] (\i+1,\k+1) -- (\i+2,\k+1) -- (\i+2,\k+2) -- (\i+1,\k+2) --(\i+1,\k+1);
		}
		}
		\node at (25/2, -0.8) {\small Carrier};
		\node[rotate=90] at (-0.8,15/2) {\small Time symbol};
		\draw[fill=TUMBeamerBlue] (26,14)--(27,14)--(27,15)--(26,15)--(26,14) node [right] at (27,14.5) {\small Pilot};
		\draw[very thin,gray] (26,12)--(27,12)--(27,13)--(26,13)--(26,12) node [right,color=black] at (27,12.5) {\small Data};
	\end{tikzpicture}
	\caption{Pilot allocation scheme of the lattice-type.}
	\label{fig:pilot_alloc}
\end{figure}
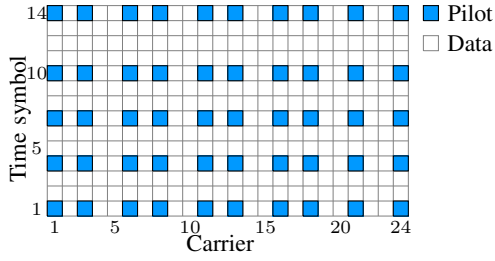

We consider a \ac{siso} transmission in the spatial domain over a doubly-selective fading channel, where $ \B H\in \C^{N_c\times N_t} $ represents the time-frequency response of the channel for $ N_c $ subcarriers and $ N_t $ time slots.
We consider a \ac{dl} \ac{fdd} system where the channel between \ac{ul} and \ac{dl} is not reciprocal, but we can exploit the recently proposed distributional shift invariance of the \ac{ul} and \ac{dl} \cite{9714227}. To this end, the training process can be performed centralized at the \ac{bs} and the parameters are offloaded to the \ac{mt}, where the \ac{dl}-\ac{ce} is performed \cite{9739092}. However, the proposed approaches are not limited to this and can be equivalently used in the \ac{ul} and in a \ac{tdd} system.

When only $ N_p $ positions of the $N_t \times N_c$ time-frequency response are occupied by pilot symbols, there is a \textit{selection matrix} $ \mbA \in \{0,1\}^{N_p\times N_c N_t} $ which represents the pilot positions.
This leads to pilot observations  
\begin{equation}\label{eq:system_model}
    \B y = \B A \B h + \B n\in\C^{N_p}
\end{equation}
with $ \mbh =\vect(\mbH)\in\C^{N_c N_t} $ and additive Gaussian noise $\B n\sim\calN_{\C}(\B 0, \B C_{\B n})$. 
In this work, we consider lattice-type pilots and the allocation scheme as shown in Fig.~\ref{fig:pilot_alloc}.

For the construction of a scenario-specific channel dataset, we use the \quadriga channel simulator \cite{QuaDRiGa1,QuaDRiGa2}.
\quadriga models the channel of the $ c $-th carrier and $t$-th time symbol as
$H_{c,t} = \sum_{\ell=1}^{L} G_{\ell} e^{-2\pi j f_c \tau_{\ell,t}}$,
where \( \ell \) is the path number and $L$ the number of multi-path components.
The frequency of the \( c \)-th carrier is denoted by \( f_c \) and the \( \ell\)-th path delay of the \( t \)-th time symbol by \( \tau_{\ell, t} \).
The coefficient \( G_{\ell} \) comprises the attenuation of a path, the antenna radiation pattern weighting, and the polarization.
We consider an \ac{uma} scenario following the 3GPP 38.901 specification, where the \ac{bs} is placed at height of 25m and covers a sector of 120°.
Each \ac{mt} is either placed indoor (80\%) or outdoor (20\%) and moves with a certain velocity $v$ in a random direction, which is captured by a drifting model.
The generated channels are post-processed to remove the path gain~\cite{QuaDRiGa2}.


\section{Power Delay Profile and Doppler Spectrum}
Practical \ac{ce} requires low complexity and memory overhead. To this end, it is beneficial to estimate the channel based on parameters that are simple to measure in practice, such as \ac{pdp} and \ac{dps} \cite{6082363,4012493}. 
Many works have considered estimating the \ac{pdp} and \ac{dps} for specific system and channel models. In this work, we assess the instantaneous \ac{pdp} and \ac{dps} by evaluating the true underlying channel realization from the \quadriga simulator. We calculate the instantaneous genie-\ac{pdp} $\B p_i$ of the $i$th time symbol $\B h_i \in\C^{N_c}$ (i.e., the $i$th column of $\B H$) 
and the genie-\ac{dps} $\B d_k$ of the $k$th carrier $\B g_k$ (i.e., the $k$th column of $\B H\T$) as
$\B p_i = |\B F_{N_c}\h\B h_i|^2$ and
$\B d_k = |\B F_{N_t}\B g_k|^2$, 
where $\B F_M$ is the $M\times M$ \ac{dft} matrix and $|\cdot|$ denotes the element-wise magnitude.
This approach is clearly utopian since the true channel that has to be estimated is used. However, in this way, we get a best-possible performance evaluation that can not be outperformed by estimating the \ac{pdp} and \ac{dps}.

Once these parameters are computed, the instantaneous frequency and time covariance matrices of the channel can be obtained due to the Wiener–Khinchin theorem \cite{89767565} as
$\B C^{\pdp}_{i} = \B F_{N_c}\diag(\B p_i) \B F_{N_c}\h$ and 
$\B C^{\dps}_{k} = \B F_{N_t}\h \diag(\B d_k) \B F_{N_t}$,
respectively,
where $\B C^{\pdp}_{i}$ and $\B C^{\dps}_{k}$ denote the frequency and time covariance matrix according to the $i$th time symbol and $k$th carrier. 
 These covariance estimates can now be used for evaluating the well-known \ac{lmmse} formula for subsequent 1D estimation of the frequency and time domain or by 2D estimation via the Kronecker product $\B C^{\pdp} \otimes \B C^{\dps}$ where the \ac{pdp} (\ac{dps}) is averaged over the $N_t$ ($N_c$) time symbols (carriers).

\section{Proposed Low-Complexity Estimators}
We briefly introduce the \acp{gmm} and the \ac{cme} based thereon from \cite{Koller2021icassp,4536367}.
A \ac{gmm} with $K$ components is a \ac{pdf} of the form
$f_{\B h}^{(K)}(\B h) = \sum_{k=1}^K p(k) \calN_{\C}(\B h; \B \mu_k, \B C_k)$
consisting of a weighted sum of $ K $ Gaussian \acp{pdf}~\cite[Sec. 9.2]{bookBi06}.
The probabilities $ p(k) $ are called \textit{mixing coefficients}, and $ \B \mu_k$ and $ \B C_k$ denote the mean vector and covariance matrix of the $ k $th \ac{gmm} component, respectively.
As explained in~\cite[Sec. 9.2]{bookBi06}, \acp{gmm} allow to calculate the \textit{responsibilities} $ p(k \mid \mbh) $ by evaluating Gaussian likelihoods.

Given data samples, an \ac{em} algorithm can be used to fit a $ K $-components \ac{gmm}~\cite[Sec. 9.2]{bookBi06}.
At this point, we want to explicitly highlight the fact that \acp{gmm} are able to approximate any continuous \ac{pdf} arbitrarily well \cite{NgNgChMc20}.
In \cite{Koller2021icassp,4536367}, a \ac{cme} is formulated based on \acp{gmm}, which is proven to asymptotically converge to the true \ac{cme} when $K$ grows large. The estimator is formulated as a weighted sum of \ac{lmmse} terms, given as
\begin{equation}\label{eq:gmm_full}
        \hat{\B h}^{(K)} = \sum_{k=1}^K p(k \mid \B y) ( \B \mu_k + \B C_k\B A\h \B C_{\B y,k}\inv (\B y - \B A\B \mu_k))
\end{equation}
where the responsibilities $p(k \mid \B y)$ are computed by 
\begin{equation}\label{eq:gmm_likelihood}
    p(k \mid \B y) = \frac{p(k) \calN_{\C}(\B y; \B A\B\mu_k, \B C_{\B y,k}) }{\sum_{i=1}^K p(i) \calN_{\C}(\B y; \B A\B \mu_i, \B C_{\B y,i}) }
\end{equation}
with $\B C_{\B y,k} = \B A \B C_k \B A\h + \B C_{\B n}$.
We denote the estimator that evaluates \eqref{eq:gmm_full} for unrestricted covariance matrices as \textit{GMM full}.

\subsection{Toeplitz Estimator}\label{sec:toep}
In the considered system model with a constant carrier-spacing and time-sampling, 
the frequency- and time-domain covariance matices can be reasonably approximated by a Toeplitz matrix, respectively. Thus, for the vectorized channel, the covariance matrix resembles a block-Toeplitz matrix with Toeplitz blocks. The class of channel covariance matrices that are expressed as
\begin{equation}\label{eq:cov_blocktoep}
    \B C = \tilde{\B Q}\h \diag(\B c) \tilde{\B Q}\in\C^{N_cN_t\times N_cN_t}
\end{equation}
are exactly the positive-definite Hermitian block-Toeplitz matrices with Toeplitz blocks, where $\tilde{\B Q}= \B Q_{N_t} \otimes \B Q_{N_c}$. Further, $\B Q_{M}$ contains the first $M$ columns of a $2M\times 2M$ \ac{dft} matrix, and $\B c\in\R_+^{4N_cN_t}$ \cite{Strang1986}.
The description of a Toeplitz-structured covariance matrix in \eqref{eq:cov_blocktoep} exactly refers to the form that is investigated in \cite{342500} for covariance estimation using the \ac{em} algorithm. That is, we can apply the derived matrix transformations in the M-step in order to have the desired structured covariances in the \ac{gmm}. In the unconstrained case, the update of the covariance matrix of the $k$th component in the M-step of the $i$th \ac{em} iteration is computed by means of the sample covariance matrix $\hat{\B C}_k^{(i)}$.
However, as suggested in \cite{342500}, if the resulting covariance matrix $\B C_k^{(i+1)}$ has the form of \eqref{eq:cov_blocktoep}, the update of the covariance matrix in the M-step is
\begin{align}
\label{eqf:inv_em}
    \B C_k^{(i+1)} &= \tilde{\B Q}\h \diag(\B c_k^{(i+1)} )\tilde{\B Q},\\
\label{eqf:inv_em1}
    \B c_k^{(i+1)} &= \B c_k^{(i)} + \diag\left(\diag(\B c_k^{(i)}) \B \Theta_k^{(i)} \diag(\B c_k^{(i)} )\right), \\
    \label{eqf:inv_em2}
    \B \Theta_k^{(i)} &= \tilde{\B Q} \left(\B C_k^{(i),-1} \hat{\B C}_k^{(i)} \B C_k^{(i),-1} - \B C_k^{(i),-1} \right)  \tilde{\B Q}\h.
\end{align}

After the fitting process, only the diagonal vectors $\B c_k$ for $k=1,\dots,K$ have to be stored. The computation of matrix-vector products with $\tilde{\B Q}$ can be performed by applying 2D \acp{fft}, yielding complexity savings as discussed later. We refer to this estimator as \textit{GMM b-toep}.

\subsection{Circulant Estimator}\label{sec:circ}
Toeplitz matrices are asymptotically equivalent to corresponding circulant matrices, cf. \cite{CIT-006}. The approximation with circulant covariance matrices is less general than the Toeplitz approximation, but allows for further simplifications.
It is well-known that the class of block-circulant matrices is diagonalized by the 2D-\ac{dft} matrix $\B F_{N_t,N_c} = \B F_{N_t}\otimes \B F_{N_c}$, i.e.,
$\B C = \B F_{N_t,N_c}\h \diag(\B c) \B F_{N_t,N_c}$,
where $\B c\in\R_+^{N_cN_t}$.
Similarly explained in \cite{4536367}, this allows for fitting a \ac{gmm} with diagonal covariances to 2D-\ac{dft} transformed data samples and thus obtaining a \ac{gmm} of the original data with circulant covariance matrices as stated above and the means via $\B F_{N_t,N_c}\h \B \mu_k$. 
Note that this procedure can only be done because the \ac{dft} matrices are unitary, which does not hold for the transformation matrix in \eqref{eq:cov_blocktoep}. Thus, the fitting with diagonal covariances in the \ac{gmm} simplifies the training compared to the transformations in \eqref{eqf:inv_em}--\eqref{eqf:inv_em2} with full matrices. We refer to this estimator as \textit{GMM~b-circ}.

\subsection{Kronecker Estimator}
If it is assumed that the covariance matrix of the time-frequency channel is separable, it can be decomposed via the Kronecker product, i.e.,
$ \B C = \B C^{\text{time}} \otimes \B C^{\text{freq}}$.
This allows for fitting two separate \acp{gmm} 
with $K_t$ (and $K_c$) components that are purely trained with the rows (columns) of the channel matrix $\B H$ for the time (frequency) covariance matrix $\B C^{\text{time}}$ ($\B C^{\text{freq}}$).
Afterwards, the high-dimensional covariance matrix is evaluated by combinatorial computation of the Kronecker products of all components to obtain the high-dimensional \ac{gmm} with $K_cK_t$ components. The same is done for the \ac{gmm} means and the mixing coefficients are evaluated by performing a single E-step of the EM algorithm. This is similarly presented in \cite{4536367} for MIMO channels and is explained here briefly for completeness. We refer to this estimator as \textit{GMM kron}.

\subsection{2$\times$1D Estimator}
As stated firstly in \cite{598897} and used later on in various applications, the 2D \ac{lmmse} filter can be decomposed into cascaded 1D filters with only marginal performance losses, yet greatly decreasing the memory and complexity overhead. 
Thereby, the ordering of estimating first in frequency or time domain is arbitrary due to linearity. Similarly as in the Kronecker estimator, two \acp{gmm} are fitted for the time and frequency domain by splitting the channel matrix into rows and columns. However, instead of computing high-dimensional covariance matrices afterwards, \ac{ce} is performed by successively evaluating the \ac{lmmse} formula for each row and column of the 2D channel matrix, respectively. The number of pilots in the time and frequency dimensions are $N_{pt}$ and $N_{pc}$, respectively, yielding $N_p = N_{pt}N_{pc}$ and $\B A = \B A_t \otimes \B A_c$ with $\B A_t\in\{0,1\}^{N_{pt}\times N_t}$ and $\B A_c\in\{0,1\}^{N_{pc}\times N_c}$.
Consequently, the system model in \eqref{eq:system_model} can be split into a time and frequency observation by replacing $\B A$ with $\B A_t$ and $\B A_c$, as well as replacing the vectorized channel with a row and column of $\B H$. After evaluating the likelihood of the time and frequency \ac{gmm}, each row and column is estimated separately via the \ac{lmmse} formula in a successive manner. 
We refer to this estimator as \textit{GMM 2$\times$1D}.
By constraining the covariances of each low-dimensional \ac{gmm} further to a Toeplitz or circulant structure, additional memory and complexity overhead can be saved. For the Toeplitz and circulant case, the same procedure as in Sec. \ref{sec:toep} and \ref{sec:circ} is adopted by replacing the 2D- by 1D-\ac{dft} matrices. We refer to these estimators as \textit{GMM 2$\times$1D-toep} and \textit{GMM 2$\times$1D-circ}.

\subsection{Memory and Complexity Analysis}
The parameters of a \ac{gmm} are determined by the covariance matrices, the means and the mixing coefficients.
The number of parameters for each variant are displayed in Table \ref{tab:num_params} and are explicitly computed for the setting in Sec. \ref{sec:sim_results} as discussed later. 
The parameter number is especially important for the \ac{fdd} case where the parameters that are learned centralized at the \ac{bs} have to be offloaded to the \acp{mt} as reasoned in Sec.~\ref{sec:system_model}.

The online complexity of the different approaches is determined by evaluating the responsibilities \eqref{eq:gmm_likelihood} and the weighted \ac{lmmse} sum \eqref{eq:gmm_full}. Because the mixture components are fixed after training, this allows for pre-computing the filters, including the inverses of $\B C_{\B y,k}$, in an offline phase for each \ac{snr}. 
It has to be noted here that it is generally not necessary to compute the whole sum in \eqref{eq:gmm_full}, but only a small fraction of it per sample, dependent on the \ac{snr}, as shown later. 
Due to the generally unstructured noise covariance $\B C_{\B n}$ and the selection matrix $\B A$, the structurally constrained cases do not directly translate into a lower online complexity. However, in the case of white noise with $\B C_{\B n} = \sigma^2\eye$, complexity can be saved by using the Sherman–Morrison formula for computing the inverse of 
$\B C_{\B y,k} = \B A \B Q\h \diag(\B c + \sigma^2\B 1)\B Q \B A\h$
that first inverts the large matrix $\B Q\h \diag(\B c + \sigma^2\B 1)\B Q$, which is simple, and afterwards finds the inverse of the submatrix \cite{Ruiz2016}. If the whole filters for each \ac{snr} value are not pre-computed, the (block-)circulant and (block-)Toeplitz approaches clearly allow for substantial complexity savings due to the \acp{fft}.
The approaches utilizing the \ac{pdp}/\ac{dps} depend on the instantaneous channel realizations and do not allow for pre-computations. 

\renewcommand{\tablename}{Table} 
\renewcommand{\thetable}{\arabic{table}} 
\begin{table}[t]
\renewcommand\tabularxcolumn[1]{m{#1}}
\begin{center}
\begin{tabular}
{ |m{0.15\columnwidth}|m{0.58\columnwidth}|m{0.12\columnwidth}|@{}m{0pt}@{} }
 \hline
 \textbf{Name} & \textbf{Parameters} & \textbf{Example}
 &\\[5pt]
 \hline
 full  &$K(\frac{1}{2}N_c^2N_t^2 + 2N_cN_t + 1)$
 & $ 7.29\cdot 10^{6}$
 &\\[5pt]
\hline
kron &$K_c(\frac{1}{2}N_c^2+2N_c + 1) + K_t(\frac{1}{2}N_t^2+2N_t + 1)$ 
&$5.78\cdot10^3$& \\[5pt]
\hline
b-toep &$K(5N_cN_t + 1) $ 
& $2.15\cdot 10^5$& \\[5pt]
\hline
b-circ &$K(2N_cN_t + 1) $ 
& $8.61\cdot 10^4$& \\[5pt]
\hline
2$\times$1D &$K_c(\frac{1}{2}N_c^2+2N_c + 1) + K_t(\frac{1}{2}N_t^2+2N_t + 1)$ 
& $3.39\cdot 10^4$& \\[5pt]
\hline
2$\times$1D-toep &$K_c(5N_c + 1) + K_t(5N_t + 1)$  
& $1.39\cdot 10^4$& \\[5pt]
\hline
2$\times$1D-circ &$K_c(2N_c + 1) + K_t(2N_t + 1)$ 
& $5.63\cdot 10^3$& \\[5pt]
\hline
\end{tabular}
\end{center}
\caption{Parameters of the GMM-based approaches. The example numbers are computed for the setting in Sec. \ref{sec:sim_results}.}
\label{tab:num_params}
\end{table}

\section{Numerical Results}\label{sec:sim_results}

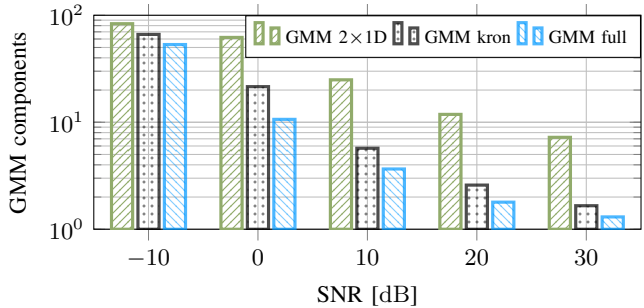
\begin{figure}[t]
	\centering
	\begin{tikzpicture}
		\begin{axis}
			[ybar,
			bar width=8pt,
			width=\plotwidth,
			height=0.5\columnwidth,
			xtick=data,
			xmin=-15, 
			xmax=35,
			xlabel={SNR [\dB]},
			ymode = log, 
			ymin= 1,
			ymax=100,
			ylabel= {GMM components}, 
			ylabel shift = 0.0cm,
			grid = both,
			legend columns = 3,
			legend entries={
			    GMM 2$\times$1D,
				\legendgmmkron,
				\legendgmm,
			},
			legend style={at={(1.0,1.0)}, anchor=north east},
			]
			\addplot[color=TUMBeamerGreen,line width=1.2pt,pattern=north east lines, pattern color=TUMBeamerGreen]
			table[x=SNR, y=nr_proba_2x1, col sep=comma]
			{csvdat/2022-04-23_17-28-55_carrier=24_symbols=14_pilots=50_pattern=lattice_sum=0.99_ntrain=100000_v=300kmh.csv};
			
			\addplot[mark options={solid},color=black,line width=1.2pt,fill=black,opacity=0.7,pattern=dots]
			table[x=SNR, y=nr_proba_kron, col sep=comma]
			{csvdat/2022-04-23_17-28-55_carrier=24_symbols=14_pilots=50_pattern=lattice_sum=0.99_ntrain=100000_v=300kmh.csv};
			
		    \addplot[mark options={solid},color=TUMBeamerBlue,line width=1.2pt,fill=TUMBeamerBlue,opacity=0.7,pattern=north west lines, pattern color=TUMBeamerBlue, style=solid]
			table[x=SNR, y=nr_proba_full, col sep=comma]
			{csvdat/2022-04-23_17-28-55_carrier=24_symbols=14_pilots=50_pattern=lattice_sum=0.99_ntrain=100000_v=300kmh.csv};
		
		\end{axis}
	\end{tikzpicture}
	\caption{Average number of GMM components necessary to achieve $99\%$ responsibility for \acp{mt} with $v\in[0;300]$km/h.}
	\label{fig:number_comp}
\end{figure}

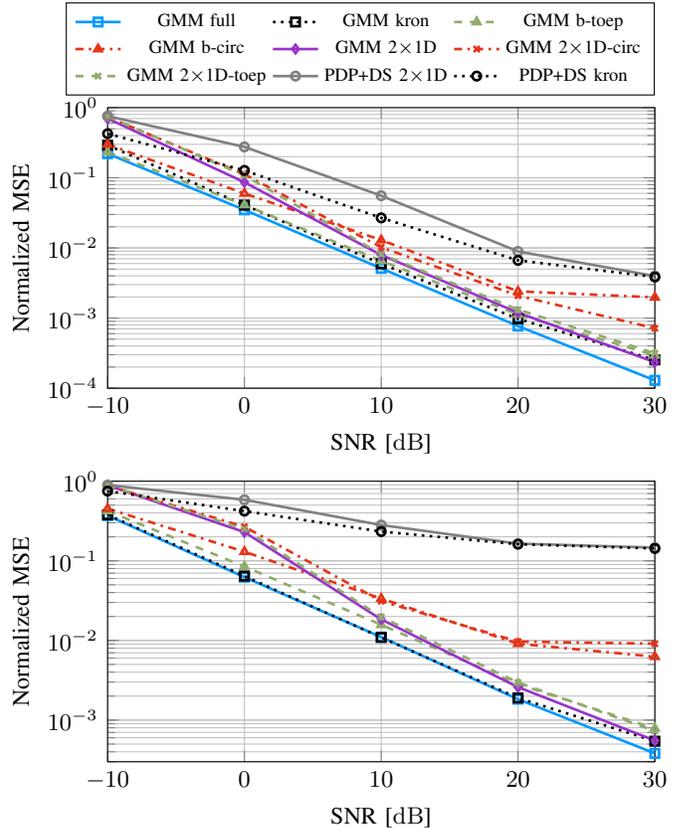
\begin{figure}[t]
\vspace{-0.14cm}
	\centering
	\begin{tikzpicture}
		\begin{axis}
			[width=\plotwidth,
			height=\plotheight,
			xtick=data,
			ymin= 1e-4,
			ymax=1,
			xmin=-10, 
			xmax=30,
			xlabel={SNR [\dB]},
			ymode = log, 
			ylabel= {Normalized MSE}, 
			ylabel shift = 0.0cm,
			grid=both,
			legend columns = 3,
			legend entries={
				\legendgmm,
				\legendgmmkron,
				GMM b-toep,
				GMM b-circ,
				\legendtwodgmmmid,
				\legendtwodgmmmidcirc,
				GMM 2$\times$1D-toep,
				\legendpdptwod,
				\legendpdpkron,
			},
			legend style={at={(1.0,1.375)}, anchor=north east, fill opacity=1, text opacity =1},
			]
			\addplot[gmm]
			table[x=SNR, y=gmm_full, col sep=comma]
			{csvdat/2022-03-29_18-25-32_carrier=24_symbols=14_pilots=50_pattern=lattice_sum=0.99_v=3kmh.csv};
		
			\addplot[gmm_kron]
			table[x=SNR, y=gmm_time_freq, col sep=comma]
			{csvdat/2022-03-29_18-25-32_carrier=24_symbols=14_pilots=50_pattern=lattice_sum=0.99_v=3kmh.csv};
			
	    	\addplot[global]
			table[x=SNR, y=gmm_2D_toep, col sep=comma]
			{csvdat/2022-04-19_09-31-30_carrier=24_symbols=14_pilots=50_pattern=lattice_sum=0.99_ntrain=100000_v=3kmh.csv};
			
			\addplot[em_gm_gamp]
			table[x=SNR, y=gmm_2D_circ, col sep=comma]
			{csvdat/2022-04-20_21-35-00_carrier=24_symbols=14_pilots=50_pattern=lattice_sum=0.99_ntrain=100000_v=3kmh.csv};
			
			\addplot[2x1gmm]
			table[x=SNR, y=gmm_2x1_freqfirst, col sep=comma]
			{csvdat/2022-03-09_23-02-43_carrier=24_symbols=14_pilots=50_pattern=lattice_v=3kmh.csv};
			
			\addplot[2x1gmm_mid_circ]
			table[x=SNR, y=gmm_2x1_freqfirst_circ, col sep=comma]
			{csvdat/2022-03-30_16-41-08_carrier=24_symbols=14_pilots=50_pattern=lattice_sum=0.99_v=3kmh.csv};
			
			\addplot[gmm_toep]
			table[x=SNR, y=gmm_2x1_freqfirst_toep_proj, col sep=comma]
			{csvdat/2022-04-05_08-30-45_carrier=24_symbols=14_pilots=50_pattern=lattice_sum=0.99_v=3kmh.csv};

			\addplot[pdp2x1]
			table[x=SNR, y=pdp_dps_freqfirst, col sep=comma]
			{csvdat/2022-03-29_18-25-32_carrier=24_symbols=14_pilots=50_pattern=lattice_sum=0.99_v=3kmh.csv};
			
			\addplot[pdp_kron]
			table[x=SNR, y=pdp_dps_kron, col sep=comma]
			{csvdat/2022-03-29_18-25-32_carrier=24_symbols=14_pilots=50_pattern=lattice_sum=0.99_v=3kmh.csv};
			
		\end{axis}
	\end{tikzpicture}
	\begin{tikzpicture}
		\begin{axis}
			[width=\plotwidth,
			height=\plotheight,
			xtick=data,
			xmin=-10, 
			xmax=30,
			xlabel={SNR [\dB]},
			ymode = log, 
			ymin= 3*1e-4,
			ymax=1,
			ylabel= {Normalized MSE}, 
			ylabel shift = 0.0cm,
			grid = both,
			]
			\addplot[gmm]
			table[x=SNR, y=gmm_full, col sep=comma]
			{csvdat/2022-03-31_17-39-40_carrier=24_symbols=14_pilots=50_pattern=lattice_sum=0.99_v=300kmh.csv};
		
			\addplot[gmm_kron]
			table[x=SNR, y=gmm_time_freq, col sep=comma]
			{csvdat/2022-03-31_17-39-40_carrier=24_symbols=14_pilots=50_pattern=lattice_sum=0.99_v=300kmh.csv};
			
			\addplot[global]
			table[x=SNR, y=gmm_2D_toep, col sep=comma]
			{csvdat/2022-04-19_09-35-23_carrier=24_symbols=14_pilots=50_pattern=lattice_sum=0.99_ntrain=100000_v=300kmh.csv};
			
			\addplot[em_gm_gamp]
			table[x=SNR, y=gmm_2D_circ, col sep=comma]
			{csvdat/2022-04-19_09-35-23_carrier=24_symbols=14_pilots=50_pattern=lattice_sum=0.99_ntrain=100000_v=300kmh.csv};
			
			\addplot[2x1gmm]
			table[x=SNR, y=gmm_2x1_freqfirst, col sep=comma]
			{csvdat/2022-03-31_17-39-40_carrier=24_symbols=14_pilots=50_pattern=lattice_sum=0.99_v=300kmh.csv};
			
			\addplot[2x1gmm_mid_circ]
			table[x=SNR, y=gmm_2x1_freqfirst_circ, col sep=comma]
			{csvdat/2022-04-01_09-21-17_carrier=24_symbols=14_pilots=50_pattern=lattice_sum=0.99_v=300kmh.csv};
			
			\addplot[gmm_toep]
			table[x=SNR, y=gmm_2x1_freqfirst_toep_proj, col sep=comma]
			{csvdat/2022-04-05_08-31-02_carrier=24_symbols=14_pilots=50_pattern=lattice_sum=0.99_v=300kmh.csv};
			
			\addplot[pdp2x1]
			table[x=SNR, y=pdp_dps_freqfirst, col sep=comma]
			{csvdat/2022-03-31_17-39-40_carrier=24_symbols=14_pilots=50_pattern=lattice_sum=0.99_v=300kmh.csv};
			
			\addplot[pdp_kron]
			table[x=SNR, y=pdp_dps_kron, col sep=comma]
			{csvdat/2022-03-31_17-39-40_carrier=24_symbols=14_pilots=50_pattern=lattice_sum=0.99_v=300kmh.csv};
			
		\end{axis}
	\end{tikzpicture}
	\caption{Channel estimation performance for \acp{mt} with $v=3$km/h (top) and $v\in[0;300]$km/h (bottom).}
	\label{fig:time_freq}
\end{figure}

\begin{figure}[t]
	\begin{center}
	\begin{tikzpicture}
		\node at (0,0) {\ref{named}};
	\end{tikzpicture}
	\end{center}
	\vspace{-0.4cm}
	\begin{subfigure}[t]{0.535\columnwidth}
		\raggedleft
		\vspace{-3.78cm}
	\begin{tikzpicture}
		\begin{axis}
			[width=1\columnwidth,
			height=0.9\columnwidth,
			xmin=1000, 
			xmax=100000,
			xlabel={\scriptsize{Training data}},
			ymode = log, 
			xmode=log,
			ymin= 1e-2,
			ymax=1e-1,
			ylabel= {\scriptsize{Normalized MSE}}, 
			ylabel shift = -0.3cm,
			grid = both,
			legend columns = 3,
			ticklabel style={font=\scriptsize},
			legend entries={
				\legendgmm,
				\legendgmmkron,
				GMM b-toep,
				GMM b-circ,
				\legendtwodgmmmid,
				 GMM 2$\times$1D-toep,
				\legendtwodgmmmidcirc,
			},
			legend to name=named,
			]
			\addplot[gmm]
			table[x=SNR, y=gmm_full, col sep=comma]
			{csvdat/2022-04-18_15-25-43_carrier=24_symbols=14_pilots=50_pattern=lattice_sum=0.99_v=300kmh_snr=10.csv};
		
			\addplot[gmm_kron]
			table[x=SNR, y=gmm_time_freq, col sep=comma]
			{csvdat/2022-04-18_15-25-43_carrier=24_symbols=14_pilots=50_pattern=lattice_sum=0.99_v=300kmh_snr=10.csv};
			
			\addplot[global]
			table[x=SNR, y=gmm_2D_toep, col sep=comma]
			{csvdat/2022-04-18_15-25-43_carrier=24_symbols=14_pilots=50_pattern=lattice_sum=0.99_v=300kmh_snr=10.csv};
			
			\addplot[em_gm_gamp]
			table[x=SNR, y=gmm_2D_circ, col sep=comma]
			{csvdat/2022-04-18_15-25-43_carrier=24_symbols=14_pilots=50_pattern=lattice_sum=0.99_v=300kmh_snr=10.csv};
			
			\addplot[2x1gmm]
			table[x=SNR, y=gmm_2x1_timefirst, col sep=comma]
			{csvdat/2022-04-18_15-25-43_carrier=24_symbols=14_pilots=50_pattern=lattice_sum=0.99_v=300kmh_snr=10.csv};
			
			\addplot[gmm_toep]
			table[x=SNR, y=gmm_2x1_timefirst_toep_proj, col sep=comma]
			{csvdat/2022-04-18_15-25-43_carrier=24_symbols=14_pilots=50_pattern=lattice_sum=0.99_v=300kmh_snr=10.csv};
			
			\addplot[2x1gmm_mid_circ]
			table[x=SNR, y=gmm_2x1_timefirst_circ, col sep=comma]
			{csvdat/2022-04-18_15-25-43_carrier=24_symbols=14_pilots=50_pattern=lattice_sum=0.99_v=300kmh_snr=10.csv};
			
		\end{axis}
	\end{tikzpicture}%
	\end{subfigure}%
	\begin{subfigure}[t]{0.535\columnwidth}
	\raggedright
	\begin{tikzpicture}
		\hspace{-0.15cm}
		\begin{axis}
			[width=1\columnwidth,
			height=0.9\columnwidth,
			xtick=data,
			xmin=8, 
			xmax=128,
			xlabel={\scriptsize{GMM components}},
			ymode = log, 
			ymin= 1e-2,
			ymax=1e-1,
			ylabel shift = 0.0cm,
			ticklabel style={font=\scriptsize},
			grid = both,
			]
			\addplot[gmm]
			table[x=comps, y=gmm_full, col sep=comma]
			{csvdat/2022-04-21_14-25-00_carrier=24_symbols=14_pilots=50_pattern=lattice_sum=0.99_ntrain=100000_v=300kmh_snr=10_comps.csv};
		
			\addplot[gmm_kron]
			table[x=comps, y=gmm_time_freq, col sep=comma]
			{csvdat/2022-04-21_14-25-00_carrier=24_symbols=14_pilots=50_pattern=lattice_sum=0.99_ntrain=100000_v=300kmh_snr=10_comps.csv};
			
			\addplot[global]
			table[x=comps, y=gmm_2D_toep, col sep=comma]
			{csvdat/2022-04-21_14-25-00_carrier=24_symbols=14_pilots=50_pattern=lattice_sum=0.99_ntrain=100000_v=300kmh_snr=10_comps.csv};
			
			\addplot[em_gm_gamp]
			table[x=comps, y=gmm_2D_circ, col sep=comma]
			{csvdat/2022-04-21_14-25-00_carrier=24_symbols=14_pilots=50_pattern=lattice_sum=0.99_ntrain=100000_v=300kmh_snr=10_comps.csv};
			
			\addplot[2x1gmm]
			table[x=comps, y=gmm_2x1_timefirst, col sep=comma]
			{csvdat/2022-04-21_14-25-00_carrier=24_symbols=14_pilots=50_pattern=lattice_sum=0.99_ntrain=100000_v=300kmh_snr=10_comps.csv};
			
			\addplot[gmm_toep]
			table[x=comps, y=gmm_2x1_timefirst_toep_proj, col sep=comma]
			{csvdat/2022-04-21_14-25-00_carrier=24_symbols=14_pilots=50_pattern=lattice_sum=0.99_ntrain=100000_v=300kmh_snr=10_comps.csv};
			
			\addplot[2x1gmm_mid_circ]
			table[x=comps, y=gmm_2x1_timefirst_circ, col sep=comma]
			{csvdat/2022-04-21_14-25-00_carrier=24_symbols=14_pilots=50_pattern=lattice_sum=0.99_ntrain=100000_v=300kmh_snr=10_comps.csv};

		\end{axis}
	\end{tikzpicture}
\end{subfigure}
	\caption{Estimation performance of the GMM-based approaches for varying numbers of training samples (left) and components (right) with $v \in [0;300]$km/h and SNR = $10$\dB.}
	\label{fig:num_data_comps}
\end{figure}
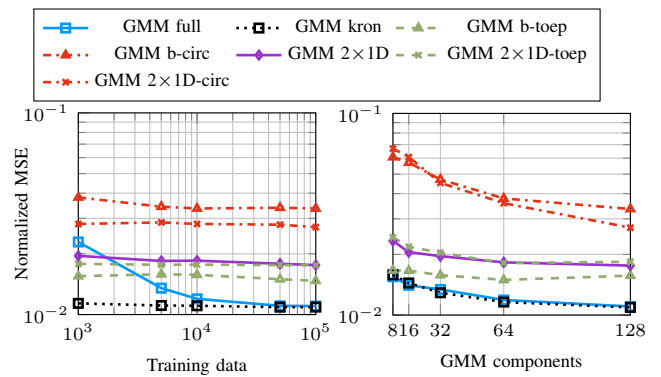

We conducted numerical experiments for the discussed system and channel model from Sec. \ref{sec:system_model} with $N_c=24$, $N_t=14$ and $N_p=50$, cf. Fig. \ref{fig:pilot_alloc}. We assume $\B C_{\B n} = \sigma^2\eye$.
The channels are normalized such that $\op E[\|\B h\|_2^2] = N_cN_t$, and the \ac{snr} is defined as $1/\sigma^2$. The MSE between the true and estimated channel is normalized by $N_cN_t$. The amount of training and test data is $10^5$ and $10^4$, respectively. The 2D \acp{gmm} have $K=128$ components. For \textit{GMM kron} and \textit{GMM 2$\times$1D}, the time and frequency domain \acp{gmm} have ($K_t=8$, $K_c=16$) and ($K_t=32,K_c=96$) components, respectively, such that their respective product and sum yield again 128 components.

For this setting, the parameters in Table \ref{tab:num_params} reveal that \textit{GMM b-toep} and \textit{GMM b-circ} only need approximately $3\%$ and $1\%$ parameters compared to \textit{GMM full}, respectively, and \textit{GMM 2$\times$1D} needs less than $0.5\%$. Most remarkably, \textit{GMM kron} and \textit{GMM 2$\times$1D-circ} need less than $0.01\%$ parameters as compared to \textit{GMM full}. 

Since $\sum_{k=1}^K p(k\mid \B y)=1$, cf. \eqref{eq:gmm_likelihood}, we evaluate the necessary mean number of \ac{gmm} components that is needed to achieve at least $99\%$ responsibility in the online estimation.
In Fig.~\ref{fig:number_comp}, it can be observed that for higher \ac{snr} values, the number of components can be reduced drastically.
A reason for this is that in low \ac{snr}, the noise variance dominates in $\B C_{\B y,k}$ from \eqref{eq:gmm_likelihood} and hence, there is higher uncertainty which component is most likely to be responsible for the underlying channel realization. Additionally, \textit{GMM kron} and \textit{GMM 2$\times$1D} need a higher number of components than \textit{GMM full}. This can be similarly reasoned, since the additional assumptions do not hold in general and thus contribute to the uncertainty. As a result, the \ac{gmm} appears to inherently perform a trade-off between more components for estimation and less parameters/complexity.

In Fig.~\ref{fig:time_freq}, we compare the \ac{gmm} variants and the \ac{pdp}/\ac{dps} based estimators with respect to their MSE performance for different \ac{snr} values. In the upper plot, we conducted simulations with \acp{mt} that have a fixed velocity of 3km/h. As expected, \textit{GMM full} yields the best performance, closely followed by \textit{GMM kron} and \textit{GMM b-toep} with negligible losses for low to medium \ac{snr} values and an increasing gap for higher \ac{snr} values. \textit{GMM b-circ} has a large gap for higher \ac{snr} values, where it saturates. \textit{GMM 2$\times$1D} has a larger gap to \textit{GMM full} in the low \ac{snr} range compared to the 2D estimators, but performs similarly well as \textit{GMM kron} for higher \ac{snr} values. Remarkably, \textit{GMM 2$\times$1D-toep} has negligible loss to \textit{GMM 2$\times$1D} over the whole \ac{snr} range, whereas \textit{GMM 2$\times$1D-circ} again shows an increasing performance gap for higher \ac{snr} values. The 2$\times$1D estimator based on \ac{pdp}/\ac{dps} is outperformed substantially by all \ac{gmm} approaches over the whole \ac{snr} range.
The Kronecker based estimator with \ac{pdp}/\ac{dps} is only better than the 2$\times$1D \ac{gmm} estimators in the low \ac{snr} range due to the higher online complexity. However, for higher \ac{snr} values it is outperformed by all \ac{gmm} variants.

In the bottom plot of Fig.~\ref{fig:time_freq}, each \ac{mt} in the cell moves with a velocity that is randomly sampled between 0--300km/h (in both the training and test data). The findings for the \ac{gmm} based estimators are in agreement with the results for a fixed $v=3$km/h. However, the performance gap of the circulant based estimators increases in the higher \ac{snr} range. The \ac{pdp}/\ac{dps} based estimators show a very poor performance with drastic losses as compared to the \ac{gmm} based approaches. To conclude, the \ac{gmm} estimators, even when having low online complexity and few parameters due to the structural constraints, outperform the estimators based on \ac{pdp} and \ac{dps} (evaluated with the ground truth channel) by far in both settings. This can be reasoned with the ability of the GMM to utilize the prior information about the scenario/environment, including side information such as the MT's mobility. 

In the left plot of Fig.~\ref{fig:num_data_comps}, we compare the \ac{gmm} variants for different numbers of training data for \acp{mt} with random velocity between 0--300km/h and \ac{snr} of 10\dB. The low-complexity approaches, due to their drastically reduced number of parameters, cf. Table \ref{tab:num_params}, are already converging for $10^3$ training samples, whereas \textit{GMM full} needs around $10^5$ training samples until convergence. This underlines the practical importance of the structurally constrained estimators.
In the right plot of Fig.~\ref{fig:num_data_comps}, we compared the \ac{gmm} approaches for the same setting as before for different number of components. \textit{GMM kron} and \textit{GMM 2$\times$1D} always yield the equivalent amount of components when multiplied or summed, respectively. As expected, the MSE decreases for higher number of components. It can be observed that the performance saturates at around 64 to 128 components, except for the (block-)circulant approaches. To summarize, increasing the number of components has the most impact on the ultra-low complexity methods
with a saturation effect for higher number of components.

\section{Conclusion}
\balance
In this work, we proposed structurally constrained \ac{gmm} based channel estimators that are designed for a specific scenario/environment. As a key result, a great amount of parameters and computational complexity can be saved with only small performance losses. A special mentioning deserves the Toeplitz based estimator that shows the best trade-off between parameters/complexity and performance. In comparison with a \ac{pdp}/\ac{dps} based estimator that is evaluated with the instantaneous ground truth data, the \ac{gmm} based approaches are clearly superior in performance due to their ability to adapt to the present environment and access this as a prior information.

\bibliographystyle{IEEEtran}
\bibliography{IEEEabrv,bibliography}

\end{document}